\documentclass[pre,twocolumn,showpacs,preprintnumbers,amsmath,amssymb]{revtex4}

\usepackage{graphicx}
\usepackage{dcolumn}
\usepackage{bm}

\newcommand{\atx}{}
\newcommand{\mll}{}
\newcommand{\ag}{}

\newcommand{\mtext}{}

\newcommand{\atext}{}
\newcommand{\mltext}{}
\newcommand{\agtext}{}
\newcommand{\mattitext}{ }
\newcommand{\newtext}{}

\newcommand{\R}{{\mathbb R}}

\def\tilde{\widetilde}
\def\bfo{\begin{eqnarray*} }
\def\efo{\end{eqnarray*} }
\def\ba{\begin{eqnarray*} }
\def\ea{\end{eqnarray*} }
\def\beq{\begin{eqnarray}}
\def\eeq{\end{eqnarray}}

\def\det {\hbox{det}}

\def\p{\partial}
\def\a{\alpha}

\def\r{\rho}

\def\p{\partial}

\def\R{\mathbb R}

\def\x{{\bf x}}
\def\y{{\bf y}}

\newcommand{\radius}{R}

\begin{document}

\title{Cloaking a Sensor via Transformation Optics}

\author{Allan Greenleaf\,${}^*$, Yaroslav Kurylev, Matti Lassas and Gunther Uhlmann}
\affiliation{Dept. of Mathematics,
 University of Rochester, Rochester, NY 14627}

\affiliation{Dept. of Mathematical Sciences, University College London,  London,
WC1E 6BT, UK}

\affiliation{Dept. of Mathematics, University of  Helsinki, FIN-00014, Finland,}

\affiliation{$\hbox{Dept. of Mathematics, University of Washington, Seattle, WA 98195}$\\
${}^*$Authors are listed in alphabetical order}



\begin{abstract}

{\atx Ideal transformation optics  cloaking at positive frequency, besides rendering the cloaked region invisible to detection by scattering of incident waves, also shields the  region from those same waves. 
In contrast, we demonstrate  that approximate cloaking permits a strong coupling  between the cloaked and uncloaked regions; }careful choice of  parameters allows this coupling to be  amplified, leading to  effective  cloaks with degraded shielding.  The sensor modes we describe are close to but distinct from interior resonances, which destroy cloaking.
As one application, we describe how to use transformation optics to hide sensors in the cloaked region and yet enable the sensors to efficiently measure  waves incident on the exterior of the cloak, an effect similar to  the plasmon based approach of Al\`u and Engheta \cite{AE}.

\end{abstract}

\maketitle

\emph{Introduction -} Transformation optics has led to designs for devices having radical effects on wave propagation,  one of the most compelling of which is cloaking \cite{PSS1,Le}. The complex material parameters of   a transformation optics cloak steer the rays around the region to be hidden, and to a large extent the behavior of the waves mimics that of the rays.  However, the  literature on such cloaks leaves the impression that  transformation optics cloaking produces a  decoupling of the waves within and external to the cloaked region, cf. \cite{AE,CummerPRE,RYNQ,McG,Cast,Gar}.  
In this  paper, we show that on the contrary,
rather than being  isolated, the cloaked region in fact  has a coupling with 
the environment surrounding the cloak, and this 
may be amplified by means of carefully chosen parameters within the cloaked region.
There is thus a fundamental difference between  cloaking for rays  and  cloaking for waves.
We emphasize that the cloaking effectiveness  can be made arbitrarily close to the ideal cloak, while keeping the sensing effectiveness fixed.
Although we focus on  scalar equations in the quasistatic and  finite frequency
regimes, modeling electrostatic \cite{GLU1} and acoustic cloaks \cite{ChenChan},  
the same phenomenon holds with regard to transformation optics  cloaking 
for transversely polarized EM waves, e.g., in the cylindrical geometry \cite{GKLU1}.
Similar considerations  allow sensors  to be cloaked from observation by static heat flow  or 
certain cases of matter waves \cite{Zhang,GKLU3}. 

For electromagnetism in the quasistatic regime,  it was already shown
in \cite{GLU1}  that  the mean voltage on the exterior surface of an ideal cloak can be measured anywhere inside the cloaked region. At nonzero frequencies,  ideal cloaking \cite{PSS1}  is accompanied by perfect {\it shielding}, meaning that an observer or device within the cloaked region cannot detect  any information  about the incident wave  \cite{GKLU1}. Furthermore,   approximate cloaking  is  also accompanied by approximate   shielding 
 \cite{RYNQ,GKLU3} except near   certain exceptional frequencies (Neumann eigenvalues of the interior), for which  the resulting resonance simultaneously destroys both the cloaking and shielding effects \nolinebreak\cite{GKLU3}.

 Here, {we consider  {\atext waves modeled by} the Helmholtz equation
 $\nabla\cdotp  \sigma(x)\nabla u(x)+\kappa(x)\omega^2u(x)=0$ {\ag in three dimensions}.
In electrostatics,  $\omega=0$ and $\sigma$ denotes  the conductivity; in acoustics,   
$\omega>0$, and $\sigma$ and $\kappa$ 
 correspond to the anisotropic mass density and  bulk modulus, resp. 
For waves governed by Helmholtz, we describe  a \emph{sensor} effect, in which the combination of an approximate cloak 
  and an inner layer implementing  a carefully chosen Robin boundary condition on its inner surface results in effective cloaking but degraded shielding, so 
that  the lowest harmonics of  incident waves penetrate deeply into the cloaked region; {\atx see Fig. 1.} 
The sensor effect occurs close to but not at interior resonances, and constitutes a new phenomenon.
{\atx In terms of the design parameters, there are three regimes: generic (standard) cloaking, resonance, and the sensor effect; see Fig. 2 (left). The double peak in Fig. 2 (right) 
at  $\alpha=  17.4$
corresponds to the deterioration  of both cloaking and the sensor effect at a resonance; the single dip to its right  at $ \alpha =  18.1$
corresponds to  the sensor mode at which cloaking is actually improved. }

\begin{figure}[htbp]
\begin{center}

\includegraphics[width=.9\linewidth]{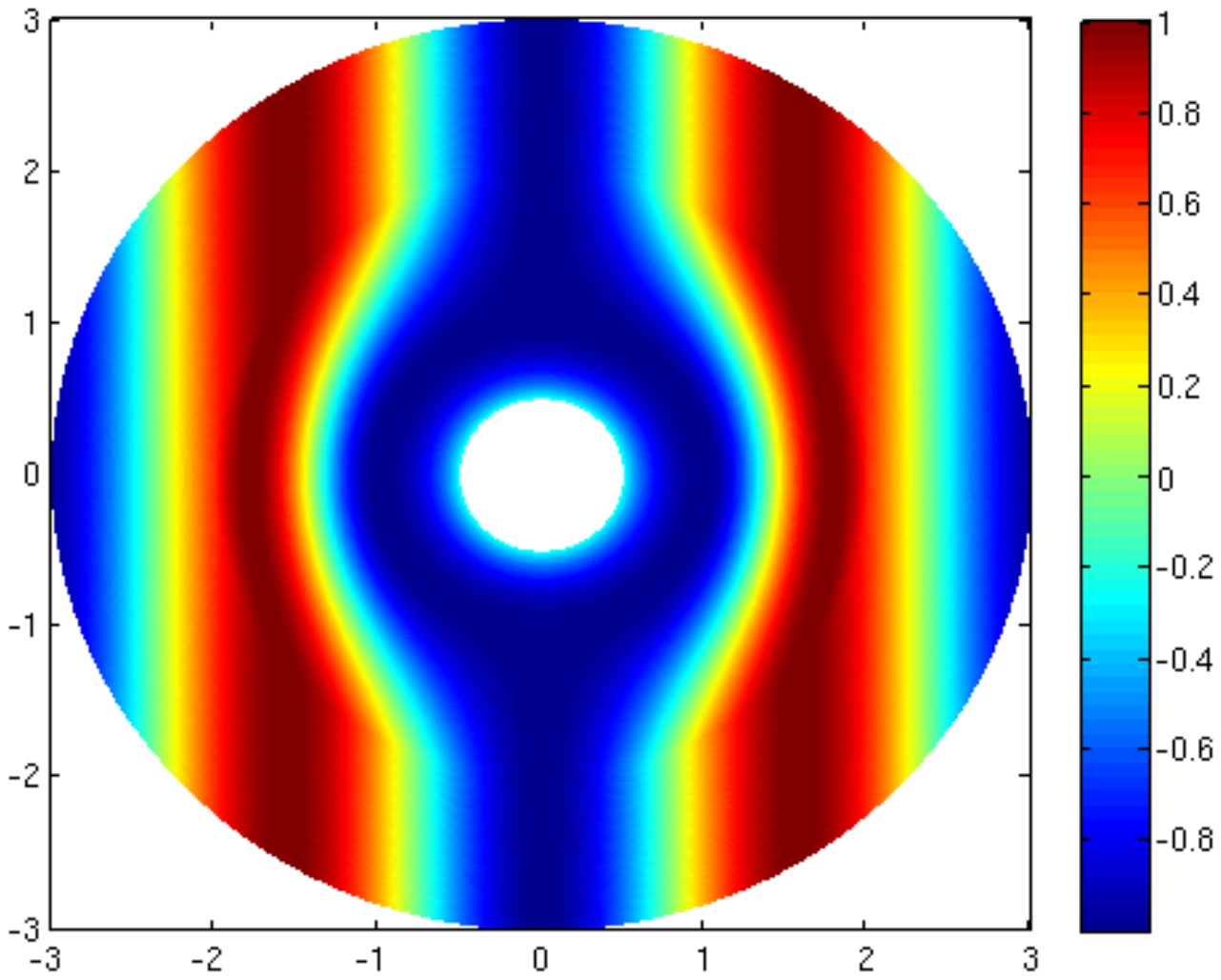}
\vspace{-2.5cm}

\includegraphics[width=1.1\linewidth]{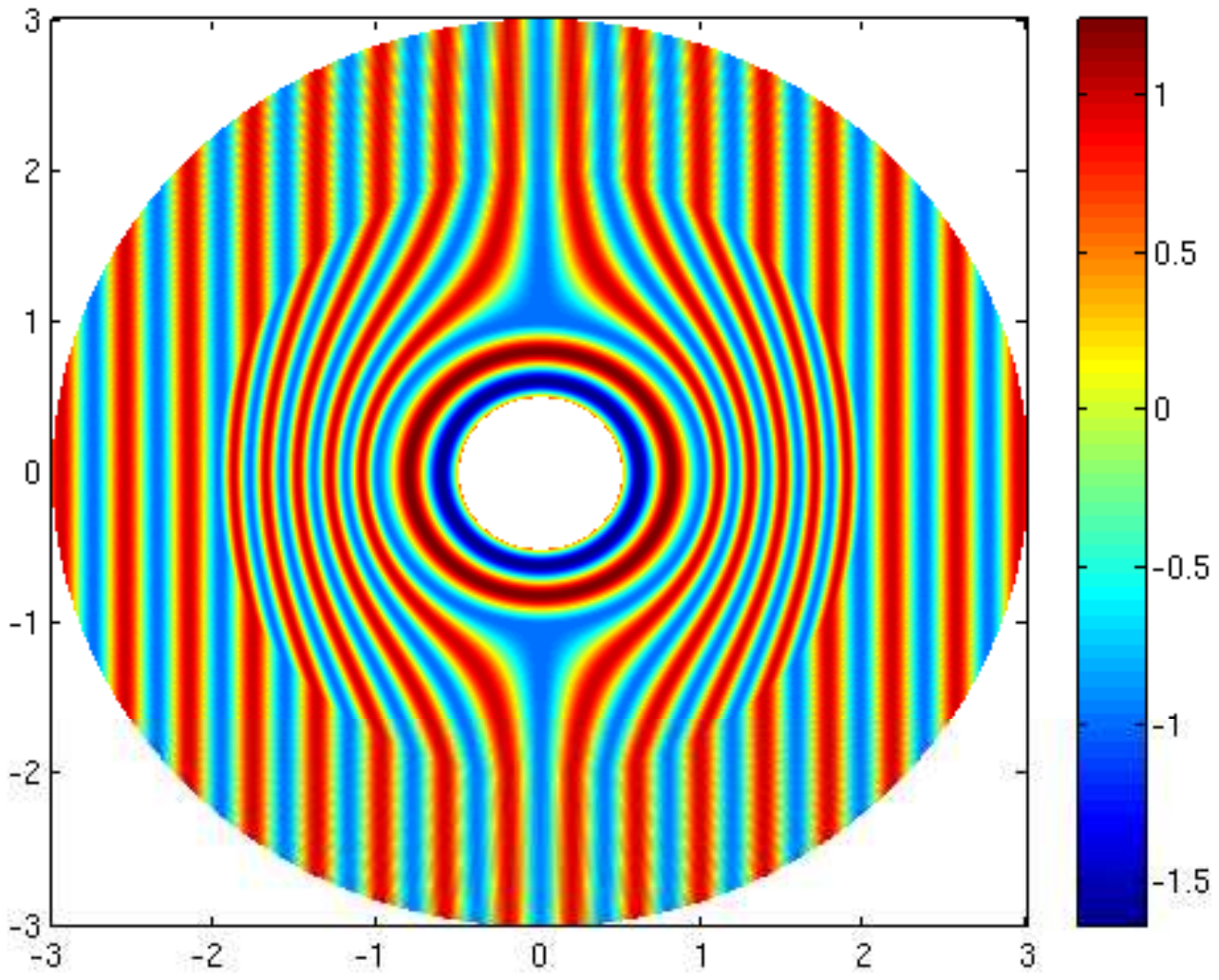}
\vspace{-3.5cm}

\end{center}
\caption{{(color online) Plane wave incident to the cloaked sensor for frequency $\omega=2$ (top) and   $\omega=16$ (bottom). See also animations \cite{vids}.
 In both figures, $\rho=10^{-2}$. Note large magnitude of fields {\atext (blue) }  and
 their radial nature in spacer layer within the cloaked region.}}
\end{figure}

In \cite{AE}, Al\`u and Engheta  showed how a plasmonic coating can be used to render an electromagnetic  sensor  almost invisible to detection by incident waves, while allowing the sensor to remain effective as a device for  measuring those very waves.  We exploit the  enhanced coupling effect described above to show that transformation optics cloaking also allows for cloaked sensors, with  a sensor embedded inside the cloaked region efficiently measuring the  incident waves without markedly altering them. 
In contrast to the plasmon based technique of \cite{AE}, this method works well for objects quite large compared with the wavelength, cf. Fig. 1(bottom). Furthermore, the quality of the approximate cloaking effect is  actually improved at a sensor mode, with the  higher order harmonics vanishing to third order in the small parameter $\r$ we use for asymptotics.  
{\mll Plasmonic cloaking is also related to the polarizability theory cf.\ \cite{Sihvola}.}

We remark that our construction, using only right-handed media,  is distinct from the anti-cloak \cite{AntiCloak}, which destroys both cloaking and shielding and requires LHM. The effect described here is also distinct from the sensitivity of the ideal cloak to small perturbations, which  degrade cloaking and shielding simultaneously \cite{RYNQ}, as well as interior resonance effects for approximate cloaks that have been reported previously \cite{GKLU3,KOVW}.
{\ag One of the surfaces in our construction is covered with material inducing a real Robin condition, corresponding to an imaginary impedance, $-i\a$; cloaking by a thin mantle of such material has been proposed by A\`u \cite{Alu}, but this appears to be a different effect than ours. }
Finally, we note that Sklan \cite{Sklan} also describes penetration of  waves into a
 cloaked region. However, the nature of that phenomenon   is
 quite different from the one described here. Indeed, \cite{Sklan} (a)  shows
 penetration of momentum into a region infinitesimally close to the cloaking surface
 (the inner surface of the cloak), corresponding to the gradient of the jump in the pressure field across the surface, while the waves we study occupy the entire cloaked region; (b) deals
 with ideal cloaking, while the phenomenon we consider occurs
 only for approximate cloaking and disappears in the ideal limit; and (c) describes effects
 which occur at any frequency, while the effects described here are
parameter sensitive, being due to a delicate energy balance near, but not at,
interior resonant frequencies.


\emph{Quasistatic regime -} For simplicity, we work throughout  with three dimensional,  spherical transformation optics  cloaking.
In the quasistatic regime, the analysis can be based on ideal cloaking for  electrostatics \cite{GLU1}. {\mltext The  solutions $u$ for the ideal  cloak considered in \cite{GLU1} are the limits of solutions for more realizable approximate cloaks.
Let $B_R$ and $ S_R$, resp.,  denote the 3D ball  of radius $R$ centered at the origin $O$, and its boundary, the 2D  sphere. For any $0<R_1<R_2$, transformation optics
(see the transformation $F_1$  in (\ref{transf}) below for $R_1=1,R_2=3$)  specifies an inhomogeneous, anisotropic cloaking conductivity $\sigma$ on the 3D spherical shell  $B_{R_2}-B_{R_1}$,  such that for  any passive object, i.e.,  one without sources or sinks  in $B_{R_1}$,  the boundary measurements of electrostatic potentials $u$ at the outer boundary $S_{R_2}$ coincide with those made when $B_{R_2}$ is filled with a homogeneous, isotropic conductivity $\sigma_0$, yielding a perfect cloaking effect. Given a boundary voltage distribution  $f$ on $S_{R_2}$, the potential $u_0$ corresponding to $\sigma_0$ takes a value at the origin, $u_0({O})=\int_{S_{R_2}} f\,dS $, where $dS$ is normalized surface measure. The potential $u$ corresponding to the same voltage $f$, but with the  object
in $B_{R_1}$ surrounded by the cloak $\sigma$, is constantly equal to $u_0(O)$ on $B_{R_1}$ \cite{GLU1}.  Thus, even {\agtext for  the ideal quasistatic cloak}, the cloaked region is \emph{not }electrically isolated, or \emph{shielded}, from its environment; rather, any passive sensor in $B_{R_1}$  measures the mean voltage at $S_{R_2}$, while not   affecting  the potential outside of $B_{R_1}$.


\emph{Finite frequency -} At finite frequency $\omega>0$, the same singular  transformation $F_1$  from virtual to physical space, applied to homogeneous, isotropic mass density $\sigma_0$ and 
bulk modulus $\kappa_0$, yields an ideal cloak corresponding to $(\sigma,\kappa)$ as in  \cite{PSS1,ChenChan}. Analysis of finite energy waves on the entire region $B_{R_2}$ was given in \cite{GKLU1}, both for the Helmholtz equation and Maxwell's equations;  we consider only the former. At frequency $\omega$, the wave $u$  within $B_{R_1}$ can be an arbitrary eigenfunction with Neumann, i.e.,  perfectly reflecting, boundary condition at $S_{R_1}$.  Thus, $u$  must vanish if $\omega$ is not {\newtext a Neumann eigenfrequency of  $B_{R_1}$, yielding an idealized decoupling of the cloaked region $B_{R_1}$ from $B_{R_2}-B_{R_1}$. This decoupling is stable, resulting in the cloaking remaining highly effective, even when the ideal cloak $(\sigma,\kappa)$  is replaced by  more physically realistic approximate cloaks,
{\newtext see  \cite{GKLU3}}. On the other hand, if $\omega$ is an eigenfrequency,   the approximate cloak supports almost trapped states, i.e., resonances, which have very small amplitude  near $S_{R_2}$ while being almost equal to  eigenfunctions of $B_{R_1}$ on $B_{R_1}$ \cite{GKLU3}.

We show below that one can create a device containing an almost cloaked yet highly sensitive sensor, consisting of  three parts: For $1<R<2$ to be chosen, we form an approximate cloak in  the spherical shell $B_2-B_{R}$, which tends toward the ideal cloak as $R\searrow 1$.  It is convenient to express parameters and asymptotics in terms of the small quantity $\r:=2(R-1)$, which $\searrow 0$ as $R\searrow 1$. Surround this approximate cloak  by  a spherical shell $B_3-B_2$ having $\sigma=\kappa=1$, on whose outer surface $S_3$ the external waves will be incident.  For a value of $R_0<1$, also to be chosen, we then embed a sensing element in the ball $B_{R_0}$,  
clad  with a surface whose ({\newtext  complex}) impedance induces a {\agtext real }Robin boundary condition, $\partial_\nu + \alpha u =0$ on $S_{R_0}$,
with $\alpha$ a {\newtext real} parameter to be chosen. 
 Finally, between the innermost  two components, we insert a   layer $B_{R} - B_{R_0}$,  
 which, for simplicity, we take to have $\sigma=\kappa=1$. 
Another variant of the above construction occurs when, instead of an obstacle $B_{R_0}$ with Robin boundary condition, 
 place inside the cloak (i)  a smaller  obstacle $B_{R_0'}$ with a Dirichlet  (sound-soft) boundary condition; and (ii)
 a spacer layer $R_0'<r<R_0$  of homogeneous and isotropic material
with $\sigma=1$ and $\kappa$ being 
a constant $\kappa_0$ chosen so that 
the radially symmetric solution of the obtained equation satisfies 
$\partial_r u(R_0) + \alpha u(R_0) =0$. 
{\ag With this Dirichlet 
obstacle and  spacer layer included,  the analysis below is valid \emph{mutatis mutandis}.
In acoustics, such a Dirichlet obstacle corresponds
to a  sound-soft surface, i.e., a freely moving boundary.
 If one places inside the ball $B(R_0')$ a device which 
measures the Neumann boundary value of the solution on 
the sound-soft boundary $\p B_{R_0'}$,
i.e., the normal component of the
moment of the surface having the
velocity $\vec V(\x,t)=\sin(\omega t)\nabla u(\x)$, then
this may be measured using another modality, say optically,
resulting in an approximately cloaked sensor which 
can measure $u_0(O)$ without significant energy loss. }

 For the device considered  above}, we first show that   $R_0$ and $\alpha=\a^{res}(\r)$  may be chosen to allow the total device to support a resonant 
 wave $\tilde u$, of very small amplitude in $B_{3} - B_{R}$ and 
 almost equal to a central (i.e., rotationally symmetric)  Robin-Neumann eigenfunction $\tilde{u}_0$ of   unit energy in the spacer
 $B_{R} - B_{R_0}$. Moreover, we show that  by varying $\a^{res}(\r)$ to a nearby value $\a^{sen}(\r)$  with
 $\a^{res}(\r)-\a^{sen}(\r)\simeq b_o\rho$ as $\rho\searrow 0$,}  the resonance effect is changed into a sensor effect:
 the resulting wave is proportional, 
 at $r=R_0$, to the value $u_0({\it O})$ that the wave would take at the origin in the absence of the device. 
 
 The  sensor is nevertheless effectively cloaked from an outside observer, since the Dirichlet-to-Neumann map on $S_{3}$,  i.e., the operator mapping the Dirichlet boundary
 value $u$ on $S_{3}$ to the Neumann boundary value $\p_\nu u$ on $S_{3}$,
 remains equal  to the free space Dirichlet-to-Neumann map, up to a perturbation order $O(\r^2)$.

\begin{figure}[htbp]
\begin{center}
\vspace{-2cm}
\hspace{-2cm}
\includegraphics[width=.7\linewidth]{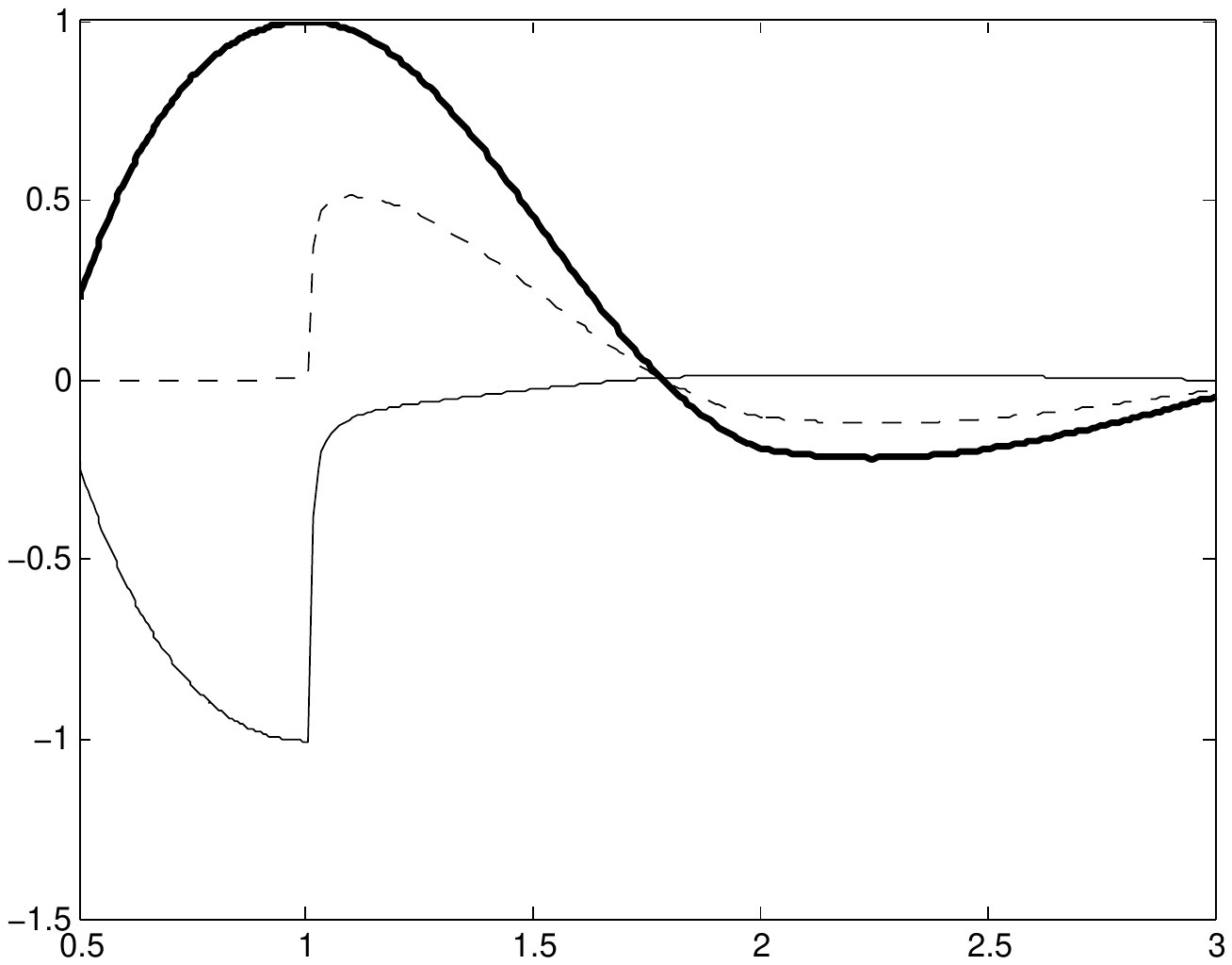}
\hspace{-2cm}
\includegraphics[width=.7\linewidth]{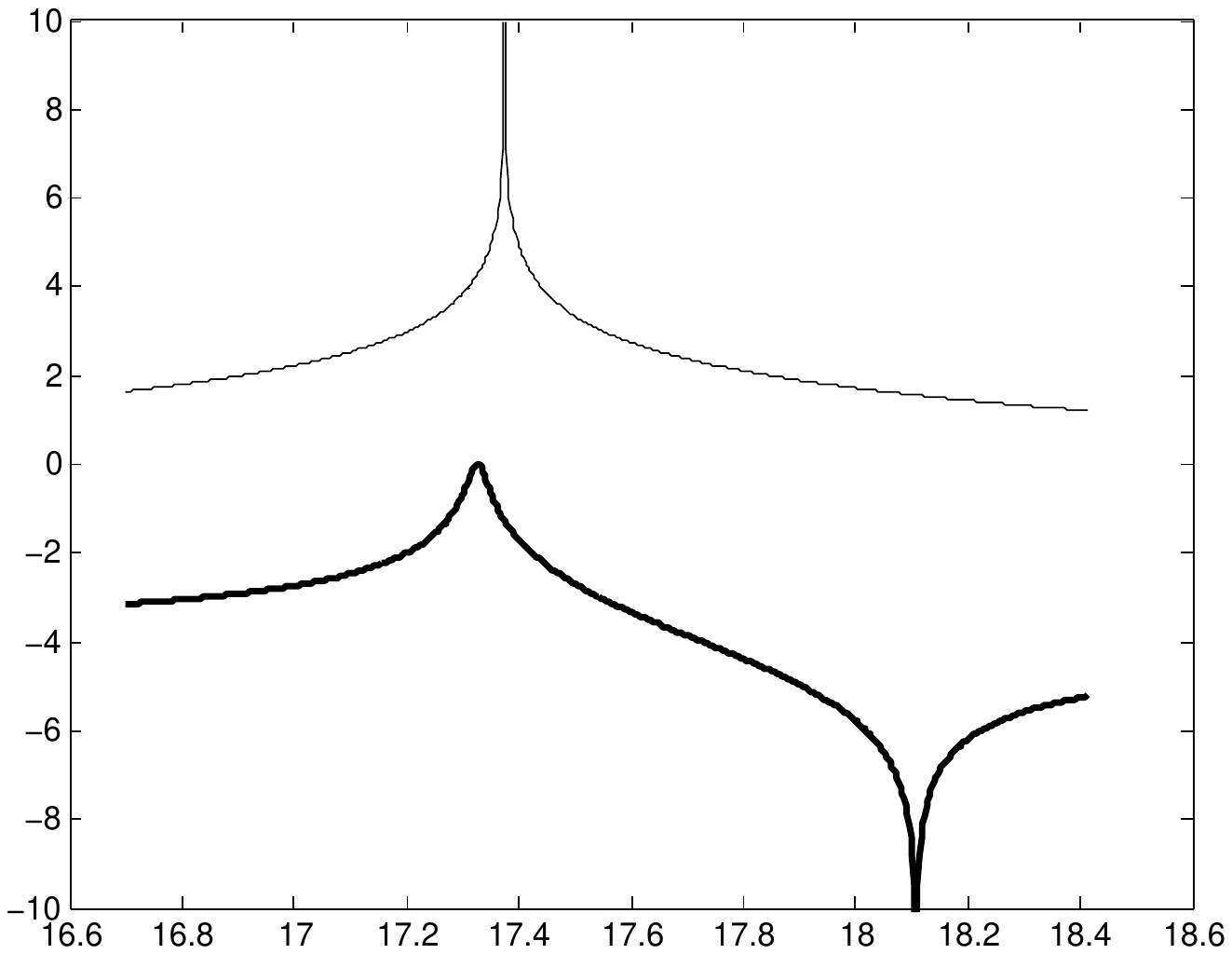}
\hspace{-2cm}
\vspace{-2cm}
\end{center}
\caption{{\newtext {\bf Left:} Radial solutions $u$ for $\a=\a^{sen}$ (thick curve), 
$\a=\a^{res}$ (thin), and for a generic $\a$  (dotted).  {\bf Right:}
{\agtext The logarithms of the sensing and scattering, $\log \Sigma$ (thin) and $\log \Gamma$ (thick), as functions of $\alpha$.  In both,
$\omega=2$ and $\rho=10^{-2}.$}}}
\end{figure}

\emph{Analysis of  approximate cloaks -} To see this in more detail, we recall some facts concerning nonsingular approximations to  ideal cloaks.
To start, let $1\le R<2$, and set $\rho=2(R-1),\, 0\le\rho<2$,  so that $R\searrow 1$ as $\rho\searrow 0$. Introduce
 the coordinate transformation 
 $F_R:B_3- B_\rho\to B_3- B_R$, 
 \beq\label{transf}
\x:=F_R(\y)=\left\{\begin{array}{cl} \y,&\hbox{for } 2< |\y|<3,\\
\left(1+\frac {|\y|}2\right)\frac{\y}{|\y|},&\hbox{for }\rho<|\y|\leq 2. 
\end{array}\right.  
\eeq
For $R=1$ ($\rho=0$),  this is the singular transformation of \cite{GLU1,PSS1}, leading to the ideal transformation optics cloak, while for $R>1$ ($\rho>0$),   $F_R$ is nonsingular  and  leads to a class of  approximate cloaks \cite{RYNQ,GKLU3,KOVW}. 
Thus,  if $\sigma_0\equiv\delta^{jk}$ denotes the homogeneous, isotropic 
tensor, then for $R>1$ the transformed tensor $\sigma^{jk}(\x)$
 is nonsingular on $R<|\x|\le 2$ , i.e., its eigenvalues are  bounded from above and below (with, however, the lower bounds of two of them going to $0$
as $R \searrow 1$).
We then define the   {\mtext approximate cloak  tensor  $\sigma_R$ on $B_3$ as
 \beq\label{grTensor}
\sigma^{jk}_R(\x)=\left\{\begin{array}{cl} \delta^{jk} ,&\hbox{for } 2< |\x|<3,\\
\sigma^{jk}(\x),&\hbox{for }R<|\y|\leq 2,\\
\delta^{jk} ,&\hbox{for } |\x|<R.
\end{array}\right.  
\eeq
Here, $\sigma(x)=(DF(x)) \sigma_0(x)\,( DF(x))^t/\det(DF(x))$ is the standard cloaking tensor.
Define also  a scalar bulk modulus,
 \beq\label{kappaRTensor}
\kappa_R(\x)=\left\{\begin{array}{cl} 1 ,&\hbox{for } 2< |\x|<3,\\
64|\x|^{-4}(|\x|-1)^4,&\hbox{for }R<|\x|\leq 2,\\
1 ,&\hbox{for } |\x|<R.
\end{array}\right.  
\eeq
}
Now place inside
the cloak a scatterer $B_{R_0}$
of radius $R_0<1$, with a surface whose impedance induces a {\agtext real} Robin  boundary condition on the sphere $S_{R_0}$. Thus, we  consider in the domain $\Omega=B_3-B_{R_0}$ the solutions of the 
problem, 
\beq\label{equat 1}
(\nabla \cdotp\sigma_R \nabla +\omega^2 \kappa_R)u_R
&=&0\quad\hbox{in }\Omega\\ \nonumber
u_R|_{S_3}=f,\,
(\p_r+\alpha)u_R|_{S_{R_0}}&=&0, 
\eeq
for an {\mattitext impedance $\lambda=-i\alpha$ to}  be specified later. As $\sigma_R$ and $\kappa_R$ are now nonsingular everywhere, 
across the internal interface  $S_R$ we have 
the standard transmission conditions, 
\beq\label{trans a1}
& & u_R|_{S_{\radius^+}}=u_R|_{S_{\radius^-}},\\  \nonumber
& &  
{\bf e_r}\cdotp \sigma_R \nabla u_R|_{S_{\radius^+}}=
 {\bf e_r}\cdotp \sigma_R \nabla u_R|_{S_{\radius^-}},
\eeq
where ${\bf e_r}$ is the radial unit vector and $\pm$ indicates  the 
trace on $S_R$ as $r\to R^\pm$.
{We note that the approximate cloak $(\sigma_R,\kappa_R)$
with anisotropic density tensor $\sigma_R$
can also be approximated, with arbitrarily precision,  by a cloak with isotropic
density \cite{GKLU3}.}

In the physical space $\Omega$  we have
 \beq \label{24.9.5}
u_R(\x)=\left\{\begin{array}{cl} v_R^+\left(F_R^{-1}(\x)\right),&\hbox{for } R<|\x|<3,\\
 v_R^-(\x),&\hbox{for } R_0<|\x|\leq R,\end{array}\right.  
\eeq
with $v_R^{\pm}$ in the virtual space, which consists of the disjoint union $(B_3-B_{\rho})\cup (B_R-B_{R_0})$,  satisfying 
\beq\label{extra-equation}
& & \hspace{-5mm}(\nabla^2+\omega^2)v_R^+(\y)=0 \quad \hbox{for
}\rho<|\y|<3,\, v_R^+|_{S_3}=f,\\
\nonumber
& &\hspace{-5mm}(\nabla^2+\omega^2)v_R^-(\y)=0, \, R_0\:<|\y|<R, 
(\p_r+\alpha)v_R^-|_{S_{R_0}}=0.
\eeq
With respect to   spherical coordinates $(r,\theta,\varphi)$, 
 the transmission conditions (\ref{trans a1})  become
$v_R^+(\rho,\theta, \phi)=v_R^-(R,\theta, \phi)$ and $\rho^2\, \p_rv_R^+(\rho,\theta, \phi)= R^2 \, \p_rv_R^-(R,\theta, \phi)$.
Since $\sigma_R,\, \kappa_R$ are spherically symmetric,
we can separate variables in (\ref{equat 1}), representing $u_R$ as
\beq \label{24.9.10}
u_R(r, \theta, \phi)= \sum_{n=0}^\infty\sum_{m=-n}^n u_R^{n,m}(r) Y_n^m(\theta, \phi),
\eeq
where $Y^m_n$ are
 the standard spherical harmonics. Then equations (\ref{equat 1}) give rise to a family
 of boundary value problems for the $u_R^{n,m}$.  For our purposes, the most important one is for the lowest harmonic term, $u_R^{0, 0}$, i.e.,  the radial component of $u_R$, which is  independent of $(\theta, \phi)$. This  is studied in the next section.

 \emph{Lowest harmonic and  the sensor effect -} 
Consider the problem (\ref{equat 1})  with $f=const=j_0(3\omega)$, 
so that  $u^\a_R(r, \theta, \phi)=u^\a_R(r)$, and  {express asymptotics of the waves  in terms of  $\r\searrow 0$.} 
 Analysis of the solutions  reveals three distinct regimes, depending on the Robin coefficient $\a$.
 This can be  seen informally, building up the wave inwards, from the exterior of the cloak, through the cloak, and into the cloaked sensor region, as follows.  {\ag On the spherical shell $\{ 2\le r\le 3\}$, a radial solution of (\ref{equat 1}) solves the ODE
\beq\label{eqn vac}
\left(r^2u'(r)\right)'+\omega^2r^2u(r)=0.
\eeq 
An incident wave has Cauchy data (i.e., the value of the wave and its normal derivative) $\left(u(3),u'(3)\right)$. Using this as initial conditions for (\ref{eqn vac})  at the right endpoint of $\{2\le r\le 3\}$, one then finds the Cauchy data at the left endpoint, $\left(u(2^+),u'(2^+)\right)$. Applying the transmission conditions  (\ref{trans a1}), one then finds $\left(u(2^-),u'(2^-)\right)$. Continuing inward through the approximate cloak, we solve the initial value problem with this Cauchy data  for 
$$\left((r-1)^2u'(r)\right)'+4\omega^2r^2u(r)=0,\, R\le r\le 2,$$
cf. \cite{GLU1,GKLU1}, evaluate the Cauchy data at $r=R^+$, and then use (\ref{trans a1}) again to find $\left(u(R^-),u'(R^-)\right)$. In the spacer layer $\{R_0\le r\le R\}$, $u$ satisfies (\ref{eqn vac}), with initial conditions at $r=R^-$ determined by the above procedure, so that finally we may evaluate $\left(u(R_0),u'(R_0)\right)$.
Starting with $\left(u(3),u'(3)\right)=(0,1)$, $\a=\a^{res}(\rho):=-u'(R_0)/u(R_0)$ gives a value of the impedance which induces a resonant, or trapped, wave, $u^{res}$. On the other hand, starting with $\left(u(3),u'(3)\right)=(j_0(3\omega),\omega j_0'(3\omega)$ gives
$\a=\a^{sen}(\rho):=-u'(R_0)/u(R_0)$  corresponding to the sensor effect and giving rise to the sensor
solution, $u^{sen}$. A calculation shows that $\a^{sen}$ is close to $\a^{res}$: for some $b_0 \neq 0$, one has
$\a^{res}(\rho)-\a^{sen}(\rho) =b_0\rho+ O(\r^2)$,  as $ \rho \searrow 0$. Thus, the sensor effect is quite sensitive to perturbations.

More careful analysis confirms the following three regimes, depending on how close $\alpha$ is to $\a^{sen}(\rho)$:
 }

 {\bf  (i)  Cloaking for generic  impedance.} If $\alpha$ is bounded away from $\alpha^{sen}$
 i.e., $|\alpha-\alpha^{sen}(\rho)|\geq c>0$, or even when
$\a=\a^{sen}(\rho)+b \rho^s$ for some $0<s<1,b\ne0$, then
the  cloak  acts as
an effective approximate cloak, and 
the field goes to zero in the cloaked region  as $\rho\searrow 0$, so that there is no  sensor effect.

{\bf (ii) Resonance effect.} For the specific value   $\a=\a^{res}(\rho)$,
  the interior resonance leads to both the  destruction  of cloaking and the absence of shielding,
since then 
 $\omega$ is {\agtext an} eigenfrequency of
the equation (\ref{equat 1}) with boundary condition $u=0$ on $\partial \Omega$
and $(\partial_r+\alpha)u=0$ on $r=R_0$.

{\bf (iii) Cloaking with sensor effect.} 
For the  value  $\a=\a^{sen}(\rho)$,
the cloak acts as an effective approximate cloak, but inside the cloaked
region the solution is proportional  to the value which the field 
would have had  at the origin in free space, with {proportionality}  {\newtext of order $O(1)$
as $\rho\to 0$, {\mltext that is, for $R_0<r<R$, 
$
u_R(r)= c^0(r) v_0+O(\r),
$
where $c^0(r)$ is a not identically vanishing function and   $v_0$ is the value at ${\it O}$ of the solution to the free-space $\nabla\cdotp\nabla v+\omega^2 v =0,$ $v|_{r=3}=f.$}
Thus,  in the sensor mode the cloak functions as an ``invisible magnifying glass",
where the value which the field would have had in the empty space  at a 
single point can be measured anywhere inside the cloak. This 
allows one to enclose 
a measurement device which does not affect the incident  fields   being measured.

If $u^{res}(r,\r)$ is the resonance solution and  $u^{cloak}(r,\r)$
is a general cloaking solutions, i.e., they  are the radial solutions of 
(\ref{extra-equation}) corresponding to $\a=\a^{res}(\rho)$ and 
a generic Robin coefficient $\a=\a^{cloak}$ bounded away from $\a^{sen}(\rho)$,
the radial sensor solution  $u^{sen}(r,\r)$ can be obtained as a linear
combination $\beta_1u^{res}(r,\r)+\beta_2u^{cloak}(r,\r)$ where $\beta_1,\beta_2\in \R$
are chosen so that the scattering vanishes, and
 $\a^{sen}(\rho)$ is obtained by computing the suitable Robin coefficient for
 this particular solution. Roughly speaking, when $\alpha=\alpha^{sen}(\rho)$,
the frequency $\omega$ is so close to the eigenfrequency of the inside of the cloak that 
 the energy flux  from the inside to outside and from the outside to inside through the
surface $r=R$  are balanced. The solution
inside the cloak does not blow up and 
the energy flux from the inside cancels the scattering caused
by the fact that the cloak is only an approximate cloak, not a perfect cloak.
Animations of the three regimes, are provided in \cite{vids}
with $\rho=10^{-2}$, $\omega=16$. For the sensor effect, $\a=-56.0$, 
 and for the generic cloaking, $\a=2$,
 we solve the scattering problem in $\R^3$ with an incident plane wave
 and use $0\leq n\leq 70$ in (\ref{24.9.10}). 
For the resonance effect, $\a=-56.1$ and 
$u_R|_{S_3}=0$.

We  now  analyze these effects quantitatively.  As a wave incident to the
cloak can be written in terms of Bessel functions $j_n(\omega r)$
and the scattered wave in terms of Hankel functions $h_n^{(1)}(\omega
r)$, the scattering effect for the zeroth order harmonic can be
numerically measured by the ratio $\Gamma= \big |\frac{c_0^{sc}}{c_0^{in}} \big |$,
where the zeroth order harmonic of the total field is
$c_0^{in} j_0(\omega r)+c_0^{sc} h_0^{(1)}(\omega r)$.
Cloaking then corresponds to $\Gamma <<1$, while the breakdown of cloaking takes place when $\Gamma=O(1)$.
Similarly,  quantify  the shielding effect 
by defining $\Sigma = {|u^{\alpha}_R(R_0) |}/ {|u^{\alpha}_R(3) |}$,
so that
$\Sigma <<1$ corresponds to effective shielding  of the cloaked region from the incident waves; $\Sigma >>1$ corresponds to the resonance regime;  and {\mll  the sensor effect corresponds to the case when
 $\Sigma$ stays between positive constants
 as $\rho \to 0$} so that} the field inside the cloaked region gives  adequate information about the external field.

A parameter $\a$ with  $|\a-\a^{res}(\r)|>c_0>0$,
or even with $|\a-\a^{res}(\r)|>c_1\rho^s,$ $c_1>0$, for some $0<s<1$,
 corresponds to the standard effects of both cloaking and shielding. The values of
$\a$  near $\a=\a^{res}$, namely
$\a-\a^{res}=O(\r^2)$, correspond to the resonance effect, with $\Sigma =O(\r^{-1})$.
{\newtext  However, a new phenomenon appears in the   region near  $\a=\a^{sen}$, namely $\a(\rho)=\a^{res}+b_0\r+O(\r^2)$.
In this region $\Gamma=O(\r^2)$ while
 $\Sigma=O(1)$, i.e., cloaking is effective but shielding is degraded. 
 Moreover, for  the higher order spherical harmonic components corresponding to $Y^m_n(\theta,\phi)$ 
 we can introduce 
figures of merit $\Gamma_{n m}(\r)$ and 
 $\Sigma_{n m}(\r)$ similar to $\Gamma,\Sigma$, resp.
 Then, for $n\ge 1$,
 \ba
 \Gamma_{n m}(\r)=O\left(\r^{(2n+1)}\right), \quad \Sigma_{n m}(\r) =O\left(\r^{(n+1)}\right),
 \ea
 showing that, for the higher harmonics there is both a strong shielding effect, i.e., lack of sensor effect, and an 
 almost ideal cloaking. 
Comparing the  formulae above with the situation for $n=0$,
 we see that the effect of the higher harmonics is 
comparatively negligible: one still has  almost ideal cloaking and sensor effects.
Hence,  if  a  measurement device  inside the obstacle $B_{R_0}$ implements this Robin boundary condition,
then the device can measure  the mean value of the wave on
 the boundary $S_{R_0}$ of the obstacle. 
  Thus}, an observer inside the cloaked region is able to
measure the outside fields, while these measurements do not disturb the field outside the cloaking device.
Observe that the cloaking and the sensing at $\a=\a^{sen}(\rho)$ is effectively independent of $\rho$, as $\rho \searrow 0$.
{\ag Recall that the cloak  in the sensor mode operates with time-harmonic
waves with given frequency $\omega$. Thus, if  a device producing a time-harmonic source 
is activated outside the cloak, there is a delay before the wave reaches its time-harmonic steady state and the zeroth order component fully
penetrate  the cloak. Since the approximate cloak corresponds to a small object
in the virtual space \cite{KOVW}, which has a small cross-section,  the smaller $\rho$ is the less energy is transferred into the cloaked region, and  the longer it takes before 
the wave reaches its sensor mode state there.
This explains why
the sensor effect disappears in the case of ideal cloaking; indeed, in
this case the time delay becomes infinite.
Finally, we again note that the fact} that $\a^{res}(\rho)-\a^{sen}(\rho)=O(\rho)$, shows that the  sensor/resonance effect
 is  highly sensitive to the value of $\a$.
 {\newtext By readjusting $\a$, an interior observer can tune this  sensor to any desired  frequency $\omega$.}

\begin{figure}[htbp]
\begin{center}
\vspace{-2.5cm}
\hspace{-2cm}
\includegraphics[width=.9\linewidth]{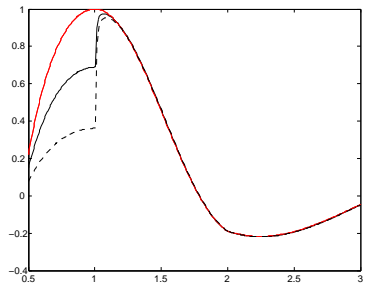}
\hspace{-2cm}
\vspace{-2.5cm}
\end{center}
\caption{(color online) Total fields corresponding to a plane wave scattering
from the  the cloaked sensor for frequency \quad$\omega=2$, $\rho=10^{-2}$,
and Robin coefficient $\a^{sen}$  (red),
$\a^{sen}+i\beta_1$  (solid black),
and $\a^{sen}+i\beta_2$ (dashed black).}
\end{figure}

\emph{Analysis of energy loss -}
 Finally, since measuring the wave on the Robin-boundary $\p B_{R_0}$ (or
 on the Dirichlet-boundary $\p B_{R_0'}$ if we use the spacer layer and
 a sound soft obstacle) will result in some energy loss, 
 we  analyze how this loss affects
 the sensing and  shielding. To this end, we compute the 
 solutions in the case when the  real Robin coefficient $\alpha$ has been replaced by a complex one, $\alpha+i\beta$. In the numerical simulations
$\alpha$ is tuned so that with the frequency $\omega=2$ and $\rho=10^{-2}$
we have sensor effect, that is, $\alpha=\alpha^{sen}(\rho)=18.1$. We considered 
 two values of $\beta$, namely when $\beta_1=3\times 10^{-2}\alpha^{sen}(\rho)$ and
$\beta_2=6\times 10^{-2} \alpha^{sen}(\rho)$, that is, the imaginary part is 
either 3\% or 6\% of the real part. Values of  $\Sigma$ (sensing)
and $\Gamma$ (scattering )  in the table below  correspond to the
solutions  in Fig.  3.
As seen in Fig. 3 and  the  table, increasing energy loss
starts to degrade the sensing effect,  but the scattering
does not grow very much. Thus, {\ag even for the parameter value
$\rho=10^{-2}$ for which the cloaking effect is relatively strong,
 the energy loss does not destroy  the cloaking property.
However, due to  energy loss the magnitude of the field becomes smaller inside
the cloak, and thus on the interior boundary. Thus,  in the presence of energy loss,
 the field needs to be measured with a more sensitive device.}

 \ba
\begin{array}{|c|c|c|}
\hline
\beta & \Sigma& \Gamma\\
\hline
0 &4.77 & 4.71\times 10^{-16}\\
\hline
\beta_1
& 3.83 & 0.0111\\
\hline
\beta_2 & 2.67 & 0.0158\\
\hline
\end{array}
\ea
\centerline{TABLE 1.  Sensing and cloaking figures of merit }
\centerline{in the presence of energy loss.}
\medskip

\emph{Discussion -} 
{We have described a new effect for transformation optics based cloaking.} In the quasistatic regime,   it was previously known that the mean voltage on the exterior surface of the cloak can be measured anywhere inside the cloaked region \cite{GLU1}. On the other hand,  at nonzero frequencies,  ideal cloaking \cite{Le,PSS1}  is accompanied by { shielding} \cite{GKLU1}:
there is a decoupling of the fields inside  and outside of the cloaked region, so that  external observations do not provide any indication of the presence of a cloaked object, nor  is any
information   about the fields outside detectable inside the cloaked region. 
Approximate   shielding goes hand in hand with 
approximate cloaking, with the shielding improving as the cloaking does,
except at resonant frequencies, at which both are destroyed \cite{GKLU3}.  
{\mtext We  have described a sensor effect  that breaks this {\atext connection} between cloaking and shielding, allowing the former without the latter, showing that transformation optics permits sensors to be cloaked. This  effect  occurs close to, but not at resonance, and is a distinct phenomenon. The sensor effect exists for any type of wave governed by (or reducible to) the Helmholtz equation, including scalar electromagnetism, acoustics and  matter waves. It would be interesting to understand to what extent it holds for other systems, such as Maxwell's equations. 
{\ag Our construction likely generalizes to Maxwell for 3D cloaks in the cylindrical geometry. However, as Maxwell's equations do not have spherically symmetric solutions,
the extension of the {sensor} effect to the  spherical geometry remains unclear.}

{\bf Acknowledgements:} AG and GU
are supported by  US NSF, YK by  UK EPSRC, ML by
CoE-213476, and GU by a Walker
 Family Endowed Professorship at UW.


\begin{thebibliography}{99}



\bibitem{AE} A. Al\`u and N. Engheta, \prl {\bf 102}, 233901 (2009).


\bibitem{Le} U. Leonhardt, Science {\bf 312}, 1777 (2006),
 U. Leonhardt and T. Tyc, Science {\bf 323}, 5910 (2008).


\bibitem{PSS1}
J.B.\ Pendry, D.\ Schurig and D.R.\ Smith,
Science  {\bf 312}, 1780  (2006).


\bibitem{CummerPRE} S. Cummer, B. Popa, D. Schurig and D. Smith,
\pre {\bf 74}, 036621 (2006).

\bibitem{RYNQ} Z.\ Ruan, M.\ Yan, C.\ Neff and M.\ Qiu, \prl {\bf  99}, 113903 (2007).

\bibitem{McG} J.\ McGuirk and P. Collins,  Opt. Exp. {\bf 16}, 17560 (2008).

\bibitem{Cast} G. Castaldi, et al., Opt. Exp. {\bf 17}, 3101 (2009).

\bibitem{Gar} F. Garc\'ia de Abajo, Physics {\bf 2}, 47 (2009).

\bibitem{GLU1}
A.\ Greenleaf, M.\ Lassas and G.\ Uhlmann, Physiol. Meas. {\bf 24}, 413 (2003); Math. Res. Lett.
{\bf 10}, 685 (2003).

\bibitem{ChenChan} H.\ Chen and  C.T.\ Chan, \apl {\bf 91}, 183518 (2007); S. Cummer, et al., \prl {\bf 100}, 024301 (2008); A.\ Greenleaf, et al, arXiv:0801.3279v1. 

\bibitem{GKLU1}
A.\ Greenleaf, Y.\ Kurylev, M.\  Lassas and G.\ Uhlmann, Comm. Math. Phys. {\bf 275}, 749 (2007).

\bibitem{Zhang} S.\ Zhang, D.\ Genov, C.\ Sun and X.\ Zhang,   \prl {\bf 100}, 123002 (2008).

\bibitem{GKLU3} A.\ Greenleaf, Y.\ Kurylev, M.\ Lassas and  G.\ Uhlmann, \prl {\bf 101},
220404 (2008); New J. Phys. {\bf 10}, 115024 (2008); arXiv:0812.1706 (2008), J. Spectral Theory, to appear.

\bibitem{Sihvola}
 I.\ Lindell,  A. Sihvola, S.\ Tretyakov, A.\ Viitanen, Electromagnetics Waves in Chiral and Bi-isotropic
Media, Artech House, 1994.

\bibitem{vids} See EPAPS Document No. [TBA] for three animated simulations of cloaking device in generic, sensor and resonance regimes.



\bibitem{AntiCloak} H. Chen, X. Luo, H. Ma and C.T. Chan, Opt. Exp. {\bf 16}, 14603 (2008).


\bibitem{KOVW} R.\ Kohn, D.\ Onofrei, M.\ Vogelius and M.\ Weinstein, Comm. Pure Appl.Math. {\bf 63},  973 (2010).

\bibitem{Alu} A. A\`u , \prb {\bf 80}, 245115 (2009).

\bibitem{Sklan} S. Sklan, \pre {\bf 81}, 016606 (2010).


\end {thebibliography}
\end{document}